\documentclass[prl,aps,twocolumn,epsfig]{revtex4}
\usepackage{graphicx}
\usepackage{dcolumn}
\usepackage{bm}

\begin{document}
\title {Criticality of tuning in athermal phase transitions}

\author{U. Chandni\footnote[1]{electronic mail:chandni@physics.iisc.ernet.in}
} \author{Arindam Ghosh}
\address{Department of Physics, Indian Institute of Science, Bangalore 560 012, India}
\author{H. S. Vijaya and S. Mohan }
\address{Department of Instrumentation, Indian Institute of Science, Bangalore 560 012, India}
\date{\today}

\begin{abstract}
We experimentally address the importance of tuning in athermal phase transitions, which are triggered only by a slowly varying external field
acting as tuning parameter. Using higher order statistics of fluctuations, a singular critical instability is detected for the first time in
spite of an apparent universal self-similar kinetics over a broad range of driving force. The results as well as the experimental technique are
likely to be of significance to many slowly driven non-equilibrium systems from geophysics to material science which display avalanche dynamics.
\end{abstract}

\maketitle

In equilibrium statistical physics, continuous phase transitions are critical and are characterized by a diverging correlation length $\xi$ at
the critical point. On the other hand, first order transitions are non-critical where the probability of phase transformation is governed by the
free energy barrier through the Boltzmann factor \cite{sornette1}. In contrast, athermal first order phase transitions are not influenced by
thermal fluctuations and proceed through a set of metastable states of free energy local minima, and hence as the external field (temperature,
magnetic field, stress etc.) is varied, display bursty and discrete avalanches \cite{reche}. Theoretical models such as random field ising model
and the renormalization group analysis \cite{sethna,sethna2} map these non-equilibrium first order phase transitions to equilibrium critical
phenomenon, although the divergence of the correlation length at the critical field has never been clearly demonstrated. Experiments are
inconclusive whether these systems self organize to the critical state over a broad range of external field
\cite{btw,zapperi,urbach,narayan,cote}, or if there exists a unique critical point that is smudged by a wide critical zone as postulated by the
concept of ``plain-old criticality'' \cite{sethna,sornette,sethna3}. The bottleneck arises since most experimental claims of critical behavior
are based on observation of a scale-free kinetics which causes power law decay in size distribution or the power spectrum of the avalanches
\cite{zapperi,cote}, but none of these are direct probes to $\xi$ itself.

In systems with many degrees of freedom \cite{nguyen,swastik}, the non-gaussian component (NGC) in  time dependent fluctuations (or noise) in
physical observables act as an indicator of long-range correlation between individual fluctuators \cite{restle,seidler}. This has been studied
in the context of Barkhausen noise from magnetization avalanches, which probes the role of long-range magnetic interactions in the domain wall
depinning when subjected to external magnetic field \cite{weissman}. Here, we focus on the avalanches in the electrical resistivity, $\rho(t)$,
during temperature-driven martensite transformation (MT), which is a prototype of athermal phase transition \cite{reche}. We demonstrate that
the NGC in avalanche induced noise is an extremely sensitive kinetic detector of criticality in continuously driven non-equilibrium systems. We
show, for the first time, the existence of a singular `global instability' ~\cite{sethna2,sornette} or divergence of $\xi$ as a function of
temperature in MT indicating, (i) mapping of non-equilibrium first order phase transition and equilibrium critical phenomena, and (ii)
conventional nature of critical behavior, even though many previous experimental results ~\cite{vives,carrillo}, as well as theoretical models
~\cite{reche2}, predict a self-organized criticality in such systems.

 The dependence of NGC on temperature ($T$) variation of $\xi$ is straightforward: For a $d$-dimensional macroscopic system of
size $L$ away from the critical regime, $\xi$ is small, and the system can be divided in $N \sim (L/\xi)^d \gg 1$ boxes so that avalanches in
each box occur independently of the others. When $N$ is large, the central limit theorem forces $\Delta\rho(t) = \rho(t) - \langle\rho\rangle$,
to be Gaussian because it is the sum of many random variables each of which represent resistivity avalanches in individual boxes
~\cite{seidler2}. ($\langle\rho\rangle$ is the time-averaged resistivity and depends only on $T$.) As $T \rightarrow T_c$, where $T_c$ is the
critical temperature, $N \rightarrow 1$ with diverging $\xi$, implying that the entire system is correlated, and $\Delta\rho(t)$ is maximally
non-Gaussian.

Our experiments were carried out with thin films of equi-atomic nickel titanium (NiTi) alloy due to the following reasons. First, Otsuka {\it et
al.} \cite{otsuka} showed that MT in NiTi is purely athermal with no detectable trace of isothermal component in avalanche dynamics. Second,
phase transitions in NiTi occur in multiple steps namely, from the high-temperature austenite phase (cubic B2: CsCl), through an intermediate
rhombohedral (R)-phase to the monoclinic B19/B19$^\prime$ martensite phase at low temperatures, which allow probing criticality in separate
ranges of $T$ in the same sweep. However, thin films of NiTi are relatively less known in terms of critical behavior, in which the disorder
component could be seriously modified by the substrate and the grain boundaries. Hence, before evaluating the NGC, we have first confirmed the
conventional signatures of avalanche dynamics and universality during MT in well-trained NiTi films with resistivity noise.

\begin{figure}
\begin{center}
\includegraphics[width=8.5cm,height=6cm]{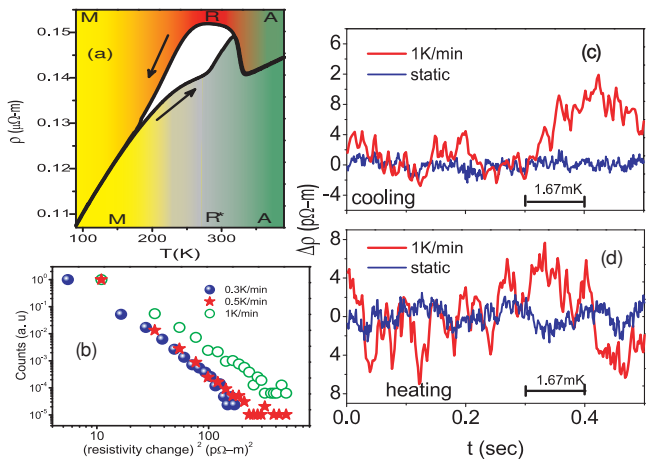}
\end{center}
\vspace{-0.7cm}\caption{(a) Resistivity vs Temperature plot for the NiTi film. Phases (M-Martensite, A-Austenite, R-Intermediate R-phase and
R*-Intermediate R*phase) during heating and cooling are shaded differently. Cooling and heating directions are indicated by arrows. (b) Size
distribution of avalanches for three different ramp rates. Resistivity fluctuations during ramping at 1K/min as well as during static noise
measurements as a function of time, for (c) cooling and (d) heating. The horizontal bar relates the change in time to that in temperature.}
\label{figure1}
\end{figure}

The $T$-dependence of $\langle\rho\rangle$ of a typical $0.9~\mu$m thick NiTi film on Si(100) substrate is shown in Fig.~1a, where each point of
$\langle\rho\rangle$ was obtained on averaging over $\approx 5$~sec. The sample was prepared by dc magnetron sputtering of a mosaic target
\cite{mohan}, which consists  of a patterned titanium disk laminated over a circular nickel disk, at an Ar pressure of $2 \times 10^{-3}$mbar
and annealed at $480^\circ$~C. Both length ($\approx 5$~mm) and width ($\approx 2$~mm) of the film were kept macroscopically large. Before
measuring noise, the system is subjected to several tens of thermal cycles till $\langle\rho\rangle - T$ traces for two successive cycles were
identical between 100~K to 370~K within the experimental accuracy. The $T$-dependence is similar to conventional bulk systems, and clearly
indicates the B2 (austenite) $\Rightarrow$ R $\Rightarrow$ B19$^\prime$ (martensite) regimes while cooling, and the reverse transformation while
heating. The transformation region can be readily identified by the hysteresis between the heating and the cooling cycles. No external stress
was applied for any of the experiments.

\begin{figure}
\begin{center}
\includegraphics[width=8.5cm,height=6cm]{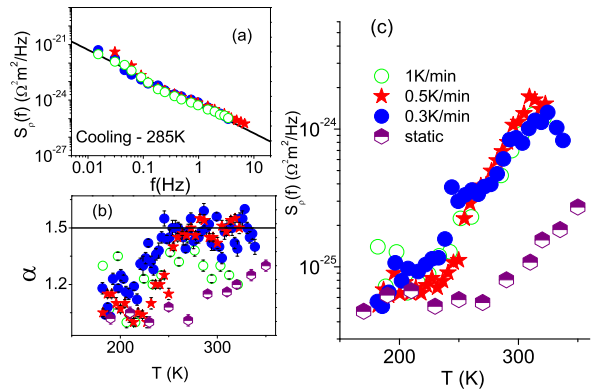}
\end{center}
\vspace{-0.7cm}\caption{(a) Power spectral density corresponding to three different cooling rates, around a mean temperature of 285K. Note that
the power spectrum is essentially independent of temperature ramping rate. (b) Frequency exponent $\alpha$ as a function of temperature, for the
different ramp rates. The clustering of points around a mean value of 1.5 indicates universality in the critical exponent. (c) Power spectral
density at 1Hz, as a function of temperature, for different ramp rates as well as during the static measurements.} \label{figure2}
\end{figure}

Time dependence of $\Delta\rho(t)$ was measured in a dynamically balanced Wheatstone bridge arrangement with an accuracy between 0.1 - 1 part
per million over a bandwidth of 16~Hz ~\cite{arindam}. Measurements were carried out in successive windows of 510 seconds while ramping $T$ at
several rates across the transition, and statistics of $\Delta\rho(t)$ were evaluated within each of these windows. Here we present the results
for three different ramp rates: 0.3~K/min, 0.5~K/min and 1.0~K/min, and also compare the results with that obtained in the static condition
where $\Delta\rho(t)$ was recorded after a 2000~sec waiting time at fixed temperatures. When $T$ was ramped, the magnitude of $\Delta\rho(t)$
was much larger during both cooling (Fig.~1c) and heating (Fig.~1d) cycles than the static case. Hence noise is dominated by avalanche dynamics
during ramping at all rates and converges to its magnitude for static case away from the hysteretic zone where no material is being transformed
(Fig.~2c). Non-zero $\Delta\rho(t)$ in the static condition indicates background activities, which could be due to the dynamic plasticity at the
parent-product interface \cite{ananth,madan}, or thermal diffusion of defects themselves ~\cite{chandni}. Fig.~1(b) shows the size distribution
function of the avalanches over the entire hysteresis region for three different rates. The avalanche or jump size can be described as the
number of mobile atoms at the parent-product interface which is roughly proportional to the square of the change between successive extrema in
$\rho$ as a function of time \cite{pelz}. Power law decay was obtained over about one and half decades with an exponent 3.1 for both 0.3K/min
and 0.5K/min and 2.7 for 1K/min. The dependence of the exponent on the driving rate is very similar to that observed in acoustic emission in
structural transitions \cite{reche3} as well as in the case of Barkhausen noise~\cite{white}, and its decrease on increasing the driving rate
can be attributed to the overlap of avalanches.

Critical behavior during MT manifests in certain universal features in the power spectral density, $S_\rho(f)$, of $\Delta\rho$, where
$S_\rho(f) \sim 1/f^\alpha$, is well-known as the $1/f$-noise \cite{sethna3,zapperi,laurson}. To illustrate this, we show $S_\rho(f)$ around $T
\approx 285$~K for cooling cycles at all three ramp rates in Fig.~2a. The absolute magnitude of $S_\rho(f)$ is nearly independent of the ramping
rate over three decades of frequency $f$ $-$ a feature that was repeated at other values of $T$ as well within the transformation zone
(Fig.~2c). This can be attributed to the purely athermal nature of MT in NiTi, so that once the disorder is quenched, the transformation
proceeds through the same set of metastable states in every thermal cycle ~\cite{reche}. This drive insensitivity is a crucial requirement for
dissipative systems to show self-similarity and universal critical exponents. Indeed, as shown in Fig.~2b, a clear clustering of the frequency
exponent $\alpha$ at $\approx 1.5\pm0.1$ for slow ramps (0.3~K/min and 0.5~K/min) over a broad range of $T$ (320~K $\rightarrow$ 240~K) can be
treated as a quantitative evidence of avalanche dynamics, as observed experimentally in vortex avalanches in superconductors ~\cite{field}, and
Barkhausen noise ~\cite{petta}, and also supported by theoretical work on interface depinning \cite{narayan} and plastic deformation
~\cite{laurson}. Both random field Ising model ~\cite{sethna3} and the two dimensional Bak-Tand-Wiesenfeld sandpile model \cite{zapperi} yield
very similar magnitude of $\alpha$ by assuming self-affine fractal avalanches within appropriate universality class. Importantly, $\alpha$
decreases, (i) to $\sim 1.30 - 1.35$ for faster ramp rate (1K/min) indicating departure from adiabatic limit ~\cite{reche3} and (ii) for
$T<240$K, probably due to dominance of smaller avalanches as the martensite fraction becomes macroscopically large.

In order to estimate the NGC in the fluctuations of $\Delta\rho(t)$, we have evaluated the second spectrum $S^{(2)}(f)$, which is the Fourier
transform of the four-point correlation function, within each window of $T$ as  $S^{(2)}(f)=\int_{0}^{\infty} \langle
\Delta\rho^{2}(t)\Delta\rho^{2}(t+\tau)\rangle_{t} \cos{(2\pi f\tau)}d\tau$. In effect, $S^{(2)}(f)$ measures a ``spectral wandering'' or
fluctuations of the power spectrum itself within a chosen frequency band ($f_L$,$f_H$), so that NGC is reflected as a non-white contribution to
the frequency-dependence of $S^{(2)}(f)$. Due to the finite detection bandwidth $f_H - f_L$, where $f_L = 1$~Hz and $f_H = 3$~Hz for our
experiments, a white Gaussian background limits the sensitivity of $S^{(2)}(f)$ to non-Gaussian effects which are hence expected to dominate
only at low frequencies ~\cite{seidler}.

\begin{figure}
\begin{center}
\includegraphics[width=7cm,height=7cm]{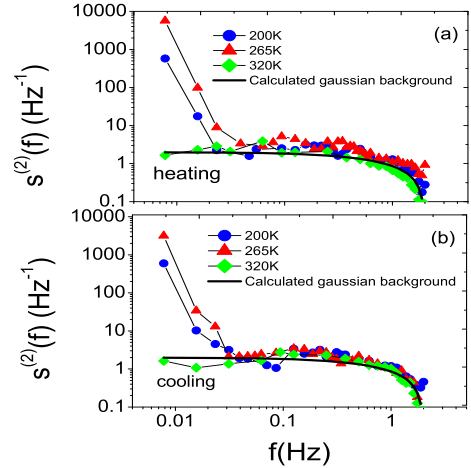}
\end{center}
\vspace{-0.7cm}\caption{Normalized second spectra for different temperatures during (a) heating and (b) cooling, at a ramp rate of 0.3K/min. The
calculated Gaussian background is plotted as well.} \label{figure3}
\end{figure}

Fig.~3 illustrates the normalized second spectrum, $s^{(2)}(f) = S^{(2)}(f)/[\int_{f_L}^{f_H}S_\rho(f)df]^2$, of $\Delta\rho(t)$ from
temperature windows of $\approx 2.5$~K centered around 200~K, 265~K and 320~K during cooling (Fig.~3a) and heating (Fig.~3b) ramps at 0.3~K/min.
The expected Gaussian background is calculated from the measured $S_\rho(f)$ within the same frequency band, and shown as the thick dark line.
While the spectrum at 320~K shows no evidence of non-Gaussian component down to $\sim 7$~mHz and agrees completely with the estimated
background, $s^{(2)}(f)$ at the other two temperatures show a steep rise at low $f$ ($\lesssim 30$~mHz). We confirmed that such a low-frequency
deviation from Gaussian background is reproduced for every thermal cycle irrespective of ramping rate (see also Fig.~4c). In order to evaluate
the total NGC contribution, here we focused on $\sigma^{(2)} = \int_0^{f_H-f_L}s^{(2)}(f)df$, and evaluated $\sigma^{(2)}$ in more than 60
successive $T$-windows over the critical regime during both heating and cooling cycles. Strikingly, $\sigma^{(2)}$ shows strong peaks during
both cooling (Fig.~4a) and heating (Fig.~4b) cycles, implying sharp increase in NGC at specific temperatures. Between the peaks, the NGC almost
vanishes and $\sigma^{(2)}$ is essentially composed of the Gaussian background (dashed line). For $T \lesssim 220$~K, the NGC increases
gradually with decreasing $T$, which we do not entirely understand at present, but it possibly indicates onset of a long-range elastic
interaction in the martensite phase. The sharp peaks in $\sigma^{(2)}$ also indicates that dynamic current redistribution is unlikely to give
rise to the observed NGC \cite{seidler2}.

\begin{figure}
\begin{center}
\includegraphics[width=8.5cm,height=6cm]{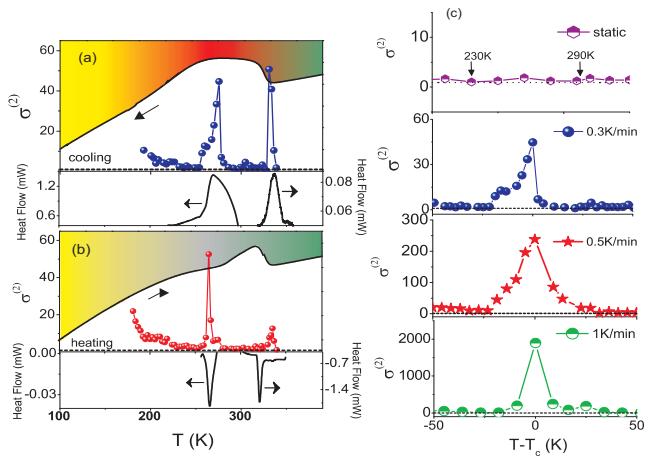}
\end{center}
\vspace{-0.7cm}\caption{Variance of the normalized second spectra as a function of temperature for (a) cooling and (b) heating, at a ramp rate
of 0.3K/min. The corresponding resistivity curves are plotted to indicate the phases. Lower panels show the corresponding differential scanning
calorimetry data as a function of temperature. (c) Variance of the normalized second spectra for different ramp rates plotted against the
deviation from the critical point ($T_c$), corresponding to the peak at the R$\Rightarrow$M transformation. Static curve is also shown to
indicate the Gaussianity of noise at long time. The calculated Gaussian background is shown as dotted lines in all the plots.} \label{figure4}
\end{figure}

The significance of the peaks became apparent when we carried out a differential scanning calorimetry(DSC) of these films over the same
temperature range. This is shown in the lower panels of Figs.~4a and 4b. Clearly, peaks in $\sigma^{(2)}$ always appear at the phase boundaries,
namely between B2$\Rightarrow$R and R$\Rightarrow$B19$^\prime$ while cooling (Fig.~4a), and at B19$^\prime$$\Rightarrow$R$^*$ and
R$^*$$\Rightarrow$B2 during the heating cycle (Fig.~4b). Hence peaks in $\sigma^{(2)}$ can be interpreted as direct manifestations of
non-Gaussianity associated with the divergence of $\xi$ during the two-stage structural phase transition in NiTi films as a function of $T$. The
sharpness of each peak establishes the requirement of tuning $T$ to the corresponding $T_c$ for the system to display ``global instability".
Fig.~4 represents the main message of this work: at the transition temperature scales defined by the DSC, which represents the latent heat
released in the first order phase transition, the correlation length diverges as well, indicating conventional equilibrium critical phenomenon.
Intriguingly, the appearance of non-Gaussianity at B2$\Rightarrow$R and R$^*$$\Rightarrow$B2 phase boundaries reveals that these transitions
contain athermal components as well.

Suppression of non-Gaussianity even slightly away from each $T_c$ implies that $\xi$ is finite. The clustering of noise frequency exponent
$\alpha$ to $\approx 1.5$ over a wide $T$ region can readily be explained by considering incoherent superposition of noise from individual boxes
of size $\xi$. From Cohn's theorem \cite{seidler2}, the measured power spectrum can be expressed as a weighted sum of noise spectrum within
individual boxes $S_{\Delta \rho}(f)=\frac{1}{I^4} \displaystyle\sum_j i_j^4 S_j (f)$, where $I$ is the total current through the sample and
$i_j$ and $S_j(f)$ refer to the current through the $j^{th}$ box and the corresponding power spectrum respectively. Away from $T_c$, the number
of boxes are large, making the second spectra gaussian, but the noise frequency dependence still reflects the universal dynamics of avalanches
within a single box.

 The rate at which the external field is driven has often been suggested as a tuning parameter which prompted us to explore $\sigma^{(2)}$ at the
R$\Rightarrow$B19$^\prime$ transition for various ramp rates in $T$. Expectedly, no NGC was observed when fluctuations were recorded at fixed
$T$ (top panel of Fig.~4c). For finite ramp rates, we did not find any perceptible change in the width of the peaks, although the magnitude of
$\sigma^{(2)}$ at the critical point increases rapidly with increasing ramp rate. This could be directly linked to the enhanced overlap of
avalanches at higher ramp rates, which increases the correlation even among avalanches well-separated in time, and thereby, causing a larger
spectral wandering ~\cite{reche3}. While the results here show singular critical points in the driving field (temperature), how sensitive is the
critical point to varying levels of disorder remains to be explored in future experiments.

In conclusion, we show for the first time, a clear existence of a singular critical point in martensite structural transition. The results
confirm a direct correspondence between non-equilibrium first order phase transition and equilibrium critical phenomenon, and at the same time
constitutes a new non-invasive technique of detecting a second order critical point that is portable to various other fields of research.

We thank Prof. Sriram Ramaswamy and Prof. Madan Rao for constructive discussions.


\begin{thebibliography}{1}
\bibitem{sornette1} D. Sornette, \emph{Critical Phenomena in Natural Sciences: Chaos, Fractals, Selforganization: Concepts and Tools }(Springer Series in Synergetics, Heidelberg,
2004).

\bibitem{reche} F. J. Perez-Reche, E. Vives, L. Manosa and A. Planes, Phys. Rev. Lett. \textbf{87}, 195701 (2001).
\bibitem{sethna2} J. P. Sethna, K. A. Dahmen and C. R. Myers, Nature \textbf{410}, 242 (2001).

\bibitem{sethna} J. P. Sethna, K. A. Dahmen, S. Kartha, J. A. Krumhansl, B. W. Roberts and J. D. Shore, Phys. Rev. Lett. \textbf{70}, 3347(1993).
\bibitem{btw} P. Bak, C. Tang and K. Wiesenfeld, Phys. Rev. Lett. \textbf{59}, 381 (1987).
\bibitem{zapperi} L. Laurson,M. J. Alava and S. Zapperi, J. Stat. Mech. \textbf{11}, L11001 (2005).
\bibitem{urbach} J. S. Urbach, R. C. Madison and J. T. Market, Phys. Rev. Lett. \textbf{75}, 276 (1995).
\bibitem{narayan} O. Narayan, Phys. Rev. Lett. \textbf{77}, 3855 (1996).
\bibitem{cote} P. J. Cote and L. V. Meisel, Phys. Rev. Lett. \textbf{67}, 1334 (1991).
\bibitem{sornette} D. Sornette, J. Phys. I. France. \textbf{4}, 209 (1994).
\bibitem{sethna3} M. C. Kuntz and J. P. Sethna, Phys. Rev. B. \textbf{62}, 11699 (2000).


\bibitem{nguyen} A. K. Nguyen and S. M. Girvin, Phys. Rev. Lett. \textbf{87}, 127205 (2001).
\bibitem{swastik} S. Kar, A. K. Raychaudhuri, A. Ghosh, H. v. Lohneysen and G. Weiss, Phys. Rev. Lett. \textbf{91}, 216603 (2003).
\bibitem{restle} P. J. Restle, R. J. Hamilton, M. B. Weissman and M. S. Love, Phys. Rev. B. \textbf{31}, 2254 (1985).
\bibitem{seidler} G. T. Seidler and S. A. Solin, Phys. Rev. B \textbf{53}, 9753 (1996).
\bibitem{weissman} K. P. O'Brien and M. B. Weissman, Phys. Rev. B. \textbf{50}, 3446 (1994).
\bibitem{vives} E. Vives, J. Ortin, L. Manosa, I. Rafols, R. Perez-Magrane and A. Planes, Phys. Rev. Lett. \textbf{72}, 1694 (1994).
\bibitem{carrillo} L. Carrillo, L. Manosa, J. Ortin, A. Planes and E. Vives, Phys. Rev. Lett. \textbf{81}, 1889 (1998).
\bibitem{reche2} F. J. Perez-Reche, L. Truskinovsky and G. Zanzotto, Phys. Rev. Lett. \textbf{99}, 075501 (2007).
\bibitem{seidler2} G. T. Seidler, S. A. Solin and A. C. Marley, Phys. Rev. Lett. \textbf{76}, 3049 (1996).
\bibitem{otsuka} K. Otsuka, X. Ren and T. Takeda, Scripta Materialia \textbf{45}, 145 (2001).
\bibitem{mohan} V. Abhilash, M. A. Sumesh and S. Mohan, Smart Mater. Struct. \textbf{14}, S323–S328 (2005).

\bibitem{arindam} A. Ghosh, S. Kar, A. Bid and A. K. Raychaudhuri, arXiv:cond-mat 0402130v1 (2004).
\bibitem{ananth} S. Sreekala and G. Ananthakrishna, Phys. Rev. Lett. \textbf{90}, 135501 (2003).
\bibitem{madan} J. Bhattacharya, A. Paul, S. Sengupta and M. Rao, arXiv:0706.3321v3 (2008).
\bibitem{chandni} U. Chandni, A. Ghosh, H. S. Vijaya and S. Mohan, Appl. Phys. Lett. \textbf{92}, 112110 (2008).
\bibitem{pelz} J. Pelz and J. Clarke, J, Phys. Rev. Lett. \textbf{55}, 738-741 (1985).
\bibitem{reche3} F. J. Perez-Reche, B. Tadic, L. Manosa, A. Planes and E. Vives, Phys. Rev. Lett. \textbf{93}, 195701 (2004).
\bibitem{white} R. A. White and K. A. Dahmen,  Phys. Rev. Lett. \textbf{91}, 085702 (2003).
\bibitem{field} S. Field, J. Witt, F. Nori and X. Ling, Phys. Rev. Lett. \textbf{74}, 1206 (1995).
\bibitem{petta} J. R. Petta, W. B. Weissman and G. Durin,  Phys. Rev. E. \textbf{57}, 6363 (1998).
\bibitem{laurson} L. Laurson and M. J. Alava, Phys. Rev. E. \textbf{74}, 066106 (2006).


\end{thebibliography}
\end{document}